\documentclass[aps,prl,twocolumn,superscriptaddress,showpacs]{revtex4-1}

\usepackage{amsmath}
\usepackage{amssymb}
\usepackage{graphicx}
\usepackage{amsbsy}

\newcommand{\BMCHI}{BMC-HI}

\newcommand{\swaprange}{{\kappa_s}} 
\newcommand{\deff}{\kappa_c} 
\newcommand{\swappingRatio}{\lambda}
\newcommand{\swappingRatioCrit}{\swappingRatio_c}
\newcommand{\swappingRatioWall}{\swappingRatio_w}
\newcommand{\density}{n}
\newcommand{\offsetInitial}{\Delta}
\newcommand{\offsetFinal}{\Delta'}
\newcommand{\vFront}{v_f}
\newcommand{\barvFront}{\bar{v}_f}
\newcommand{\correlationL}{l}

\newcommand{\sign}[1]{\textrm{sign}(#1)}

\newcommand{\halff}{{\textstyle\frac{1}{2}}}

\newcommand{\lb}{\left (}
\newcommand{\rb}{\right )}

\newcommand{\refeq}[1]{(\ref{#1})}

\begin{document}

\title{Layering instability in a confined suspension flow}

\date{\today}
\author{M.\ Zurita-Gotor}
\affiliation{Departmento de Ingenier\'ia Aeroespacial y Mec\'anica de Fluidos, Universidad de Sevilla, Sevilla 41092, Spain}
\author{J.\ B{\l}awzdziewicz}
\affiliation{Department of Mechanical Engineering, Texas Tech University, Lubbock, Texas 79409, USA}
\author{E.\ Wajnryb}
\affiliation{Institute of Fundamental Technological Research, Warsaw, Polish
Academy of Sciences}

\date{\today}

\begin{abstract}

We have shown [J.\ Fluid Mech.\ {\bf 592}, 447 (2007)] that swapping
(reversing) trajectories in confined suspension flows prevent
collisions between particles approaching each other in adjacent
streamlines.  Here we demonstrate that by inducing layering this
hydrodynamic mechanism changes the microstructure of suspensions in a
confined Couette flow.  Layers occur either in the near-wall regions
or span the whole channel width, depending on the strength of the
swapping-trajectory effect.  While our theory focuses on dilute
suspensions, we postulate that this new hydrodynamic mechanism
controls formation of a layered microstructure in such flows in a wide
range of densities.
\end{abstract}

\pacs{xxxx}

\maketitle

Confined particulate flows are important due to their applications in
microfabrication, microfluidics and biotechnology.  Recent studies
have shown complex nonlinear microstructural evolution
\cite{Thorsen-Roberts-Arnold-Quake:2001,*Cohen-Mason-Weitz:2004,
Komnik-Harting-Herrmann:2004,*Yeo-Maxey:2010c,*Xu-Rice-Dinner-Cheng-Cohen:2011,
Cui-Diamant-Lin-Rice:2004,
Beatus-Tlusty-Bar-Ziv:2006,*Beatus-Bar-Ziv-Tlusty:2007, 
Baron-Blawzdziewicz-Wajnryb:2008,
Eral-vandenEnde-Mugele-Duits:2009, 
Blawzdziewicz-Goodman-Khurana-Wajnryb-Young:2010},
strikingly different than the behavior with no confining walls.

Confinement affects the system behavior both by imposing geometrical
constraints on particle motion and by giving rise to purely
hydrodynamic mechanisms.  While geometry-driven phenomena have been
extensively studied (e.g., ordered structures in dense suspensions
and emulsions tightly confined by bounding walls
\cite{Thorsen-Roberts-Arnold-Quake:2001,*Cohen-Mason-Weitz:2004}), 
hydrodynamic mechanisms that are wall-induced have only recently been at the
center of attention.

So far the most thoroughly investigated hydrodynamic confinement
phenomenon is the fluid backflow produced by particle motion. It has
been studied in linear conduits
\cite{Cui-Diamant-Lin:2002,Navardi-Bhattacharya:2010},
and in parallel-wall channels 
\cite{Cui-Diamant-Lin-Rice:2004,Bhattacharya-Blawzdziewicz-Wajnryb:2005}.
The backflow gives rise to the anomalous sign of the mutual diffusion
coefficient for Brownian spheres
\cite{Cui-Diamant-Lin-Rice:2004,Bhattacharya-Blawzdziewicz-Wajnryb:2005},
wave propagation in flow-driven drop or particle trains
\cite{Beatus-Tlusty-Bar-Ziv:2006,*Beatus-Bar-Ziv-Tlusty:2007,
Baron-Blawzdziewicz-Wajnryb:2008,
Blawzdziewicz-Goodman-Khurana-Wajnryb-Young:2010}, 
and unusual stability of a square particle lattice
\cite{Baron-Blawzdziewicz-Wajnryb:2008,%
  Blawzdziewicz-Goodman-Khurana-Wajnryb-Young:2010}.  

\begin{figure}
  \includegraphics[width=.47\textwidth]{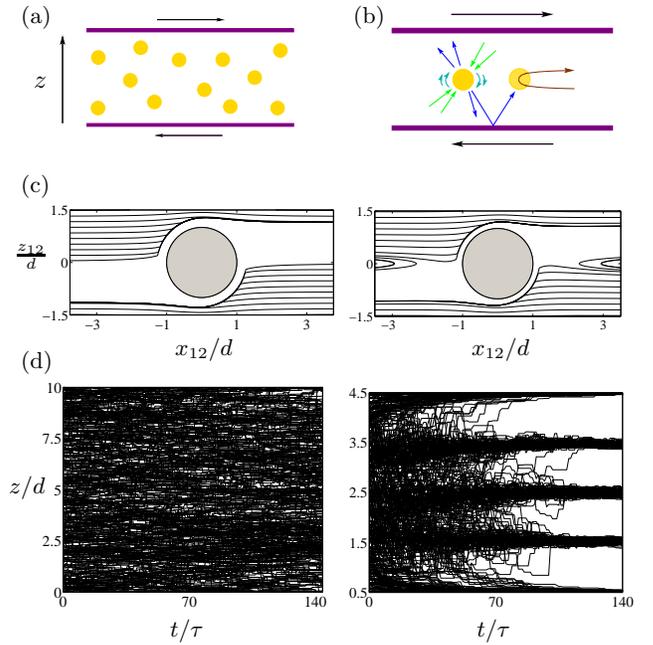}
\caption{Wall-induced layering.  (a) Suspension geometry. (b)
  Swapping-trajectory (ST) effect: the flow scattered from the walls
  produces lift causing the reversal of particle motion.  (c) Relative
  particle trajectories in two systems: unconfined (left) and confined
  with an ST region (right).  (d) Time evolution of the transverse
  position $z$ in suspension: unconfined (left) and confined (right).
  Time is normalized by the characteristic time between collisions
  $\tau=1/(n_0\dot\gamma d^3)$.  }
\label{trajectory plot version 2}
\end{figure}

Here we demonstrate that the evolution of suspension microstructure
also depends on the swapping (reversing) trajectory effect
\cite{Zurita_Gotor-Blawzdziewicz-Wajnryb:2007b,
Bossis-Meunier-Sherwood:1991}.
Because this proposed confinement-induced hydrodynamic mechanism
prevents collisions of particles in adjacent streamlines in a confined
Couette flow, a uniform suspension microstructure is destabilized,
leading to the formation of particle layers parallel to the walls
(cf., Figs.\ \ref{trajectory plot version 2} and \ref{H15-dens version
2}).

\begin{figure}
    \includegraphics[width=.47\textwidth]{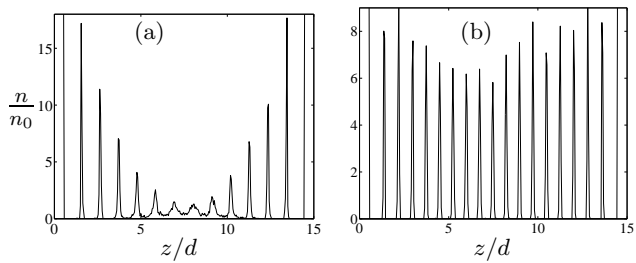}

    \caption{Suspension microstructure for particles of different
      roughness in a channel of width $H/d=15$ at time $t/\tau=2400$.
      Density profiles $\density / \density_0$ are shown vs transverse
      position $z/d$.  Roughness range (a) $\epsilon/d=0.25$; (b)
      $\epsilon/d=0.064$.}
\label{H15-dens version 2}
\end{figure}

Layered suspension microstructure that spontaneously arises in
confined flows, even at moderate particle concentrations, has recently
been observed experimentally \cite{Cohen-Cheng:2011} and in computer
simulations
\cite{Komnik-Harting-Herrmann:2004,*Yeo-Maxey:2010c,*Xu-Rice-Dinner-Cheng-Cohen:2011} 
(also in movies A1--A4 in \cite{supplemental}).  A key hydrodynamic
mechanism responsible for this behavior is revealed by our present
work.

We consider a suspension of non-Brownian spherical particles in planar
Couette flow of shear rate $\dot\gamma$ [cf., Fig.\ \ref{trajectory
    plot version 2}(a)], under creeping-flow conditions.  The
particles have finite roughness, which is modeled by a very steep,
short-range repulsive potential of the range $d_r=d+\epsilon$ (where
$d$ is the hydrodynamic diameter) to mimic contacts between roughness
asperities.  These direct particle contacts remove the Stokes-flow
symmetry of binary collisions, producing finite transverse particle
displacements \cite{daCunha-Hinch:1996}, as illustrated in
Fig.\ \ref{trajectory plot version 2}(c).

Our analysis is focused on the dilute-suspension limit. The
simulations in Figs.\ \ref{trajectory plot version 2} and
\ref{H15-dens version 2} are performed using the Boltzmann--Monte
Carlo (BMC) method, in which the system dynamics is modeled as a
sequence of uncorrelated binary collisions
\cite{Zurita_Gotor-Blawzdziewicz-Wajnryb:2007b}.  Since particle
correlations are important only at higher concentrations, this
approach is appropriate in the low-concentration regime.  We analyze
suspensions at low concentrations $n_0$ to emphasize physical
mechanisms involved in the formation of the layered microstructure.
However, our direct numerical simulations (cf., movies B1-B4 in
\cite{supplemental}) demonstrate that these mechanisms are present
also at higher concentrations.

\begin{figure}[b]
    \includegraphics[width=.45\textwidth]{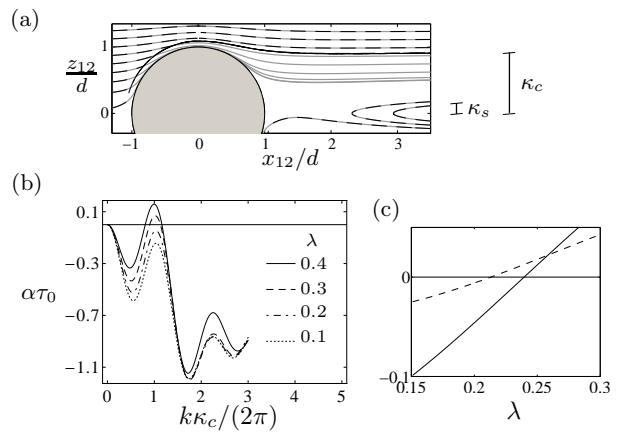}
    \caption{(a) Pair collisions: relative trajectories for roughness
      range $\epsilon/d=9 \cdot 10^{-2}$ (dashed lines) and
      $\epsilon/d=10^{-3}$ (gray solid). (b) Growth rate $\alpha$ of
      small harmonic perturbations vs the wavevector $k$ for several
      values of the swapping ratio $\lambda$ (as labeled) in model
      M2. (c) Peak value of $\alpha$ vs swapping ratio $\lambda$ for
      model M2 (solid) and M3 (dashed).  The growth-rate ratio
      $\alpha$ is normalized by the characteristic time
      $\tau_0^{-1}=\dot{\gamma} n_0 \deff^D$, where $D$ is the
      dimensionality of the model.}
\label{estability}
\end{figure}

\begin{figure}
    \includegraphics[width=.445\textwidth]{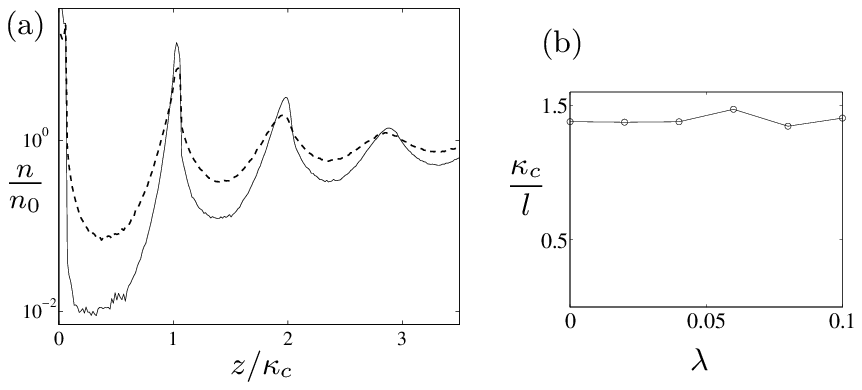}
  \caption{Long-time near-wall suspension microstructure for subcritical
  values of swapping ratio $\lambda$ (model M3) (a) Density profile
  $n/n_0$ for $\swappingRatio=0$ (dashed line) and
  $\swappingRatio=0.1$ (solid).  (b) inverse correlation length
  $l^{-1}$ vs $\swappingRatio$.}
\label{near-wall behavior stable}
\end{figure}

\begin{figure}
    \includegraphics[width=.455\textwidth]{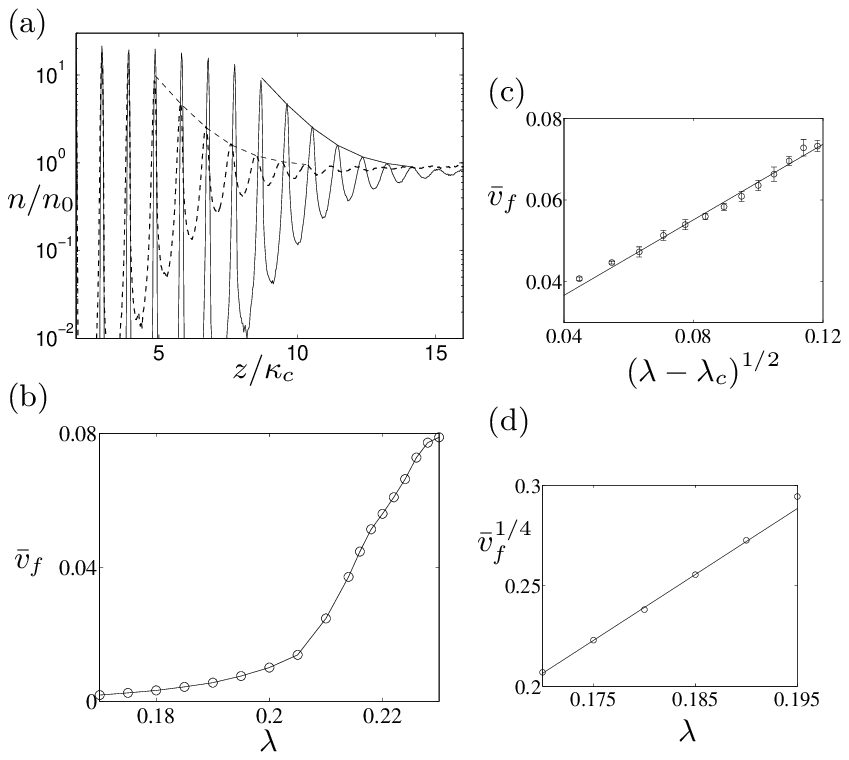}
 \caption{ Near-wall suspension microstructure for supercritical
   values of swapping ratio $\lambda$ (model M3). (a) Density profile
   at times $t_1/\tau_0=1250$ (dashed line) and $t_2/\tau_0=2500$
   (solid line) for swapping ratio $\swappingRatio=0.18$ and channel
   width $H/d=60$.  Characteristic time $\tau_0= 1/ \left ( n_0
   \dot{\gamma} \deff^3 \right )$. (b) Normalized propagation velocity
   $\barvFront=v_f \tau_0/\deff$. Near critical scaling behavior of
   $\barvFront$ for two cases: (c)
   $\swappingRatio\gtrsim\swappingRatioCrit$, and (d)
   $\swappingRatio\gtrsim\swappingRatioWall$.
\label{near-wall behavior unstable}}
\end{figure}

\begin{figure}
    \includegraphics[width=.46\textwidth]{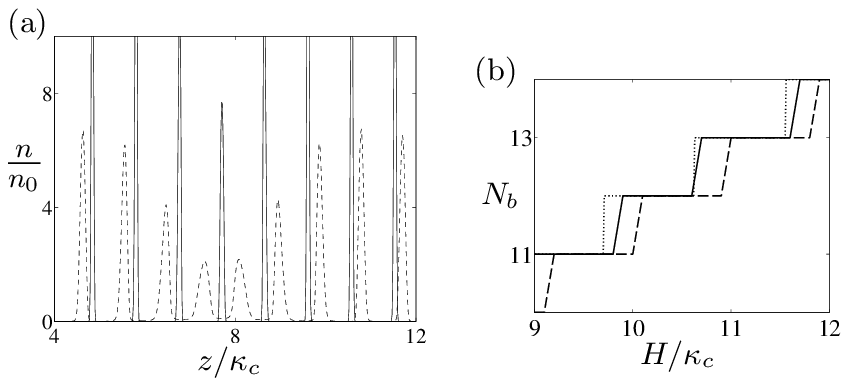}
 
\caption{Layer restructuring for supercritical swapping ratio
  $\swappingRatio=0.24$ (model M3).  (a) Merging of two central layers
  when the width is incommensurate $H/\deff=15.4$; intermediate time
  (dashed line), long time (solid). (b) Number of layers vs normalized
  channel width: maximal number during evolution (solid); steady state
  (dashed); estimate based on the wavelength of the most unstable
  Fourier mode (dotted).}
\label{commensurability}
\end{figure}

In the BMC method, the particle distribution is evolved by performing
a sequence of uncorrelated collisional steps.  In each step, we choose
a random pair of particles, simulate their binary collision, and
update the cross-streamline particle positions according to the
post-collisional displacements \cite{supplemental}.  When binary
trajectories are accurately evaluated taking into account the
hydrodynamic interactions (HI) in the wall presence
\cite{Bhattacharya-Blawzdziewicz-Wajnryb:2005,supplemental}, we refer
to this description as \BMCHI.  To highlight the role of topological
features of pair trajectories without incurring large numerical cost
of evaluation of HI, we use two simplified collision models M2 and M3
(described below).  Figures \ref{trajectory plot version 2} and
\ref{H15-dens version 2} were obtained using BMC-HI, and
Figs.\ \ref{estability}--\ref{commensurability} correspond to models
M2 and M3.

Figure \ref{trajectory plot version 2}(d) demonstrates that a
suspension confined between two parallel walls develops a well-defined
layered structure (after about 20 collisions per particle), while the
unconfined system remains uniform.  Fig.\ \ref{H15-dens version 2}
illustrates a strong dependence of the layered structure on the range
$\epsilon$ of non-hydrodynamic interparticle interactions.

We argue that the observed layering behavior stems from the
swapping-trajectory (ST) effect that causes approaching particles to
reverse their motion, and avoid collision.  Such trajectories, which
occur in the wall presence [cf., Fig.\ \ref{trajectory plot version
2}(c)], prevent large collisional displacements for particles with a
sufficiently small transverse offset, leading, therefore, to
stabilization of particle layers.

The ST domain results from wall-mediated HI
between the approaching particles
\cite{Zurita_Gotor-Blawzdziewicz-Wajnryb:2007b}. 
The wall reflection of the perturbation flow produced by one of the
particles [cf., Fig.\ \ref{trajectory plot version 2}(b)] pushes the
other particle across streamlines of the applied flow toward the fluid
moving in the opposite direction, causing the reversal of the relative
particle motion.  Through this purely hydrodynamic mechanism confining
walls influence the suspension microstructure.

\paragraph*{Population-balance equation --}  
To elucidate the effect of swapping trajectories on the suspension
dynamics we use the population-balance method, where the particle
density $\density(z)$ (uniform in the flow and vorticity directions
$x$ and $y$) is described by a master equation that accounts for the
effect of binary particle collisions on the suspension motion.  In the
simplest two-dimensional (2D) version (with no $y$ direction), the
master equation reads
\begin{eqnarray}
    \label{population-balance equation}
    \frac{\partial \density \lb z,t \rb }{\partial t}&=&
    \dot\gamma\int_{-\infty}^\infty 
    [
    \density(z+\halff\offsetInitial-\halff\offsetFinal)
    \density(z-\halff\offsetInitial-\halff\offsetFinal)
    \nonumber\\%
    &-&\density(z)\density(z-\offsetInitial)
    ]\,\offsetInitial\,d\offsetInitial,
\end{eqnarray}
where $\offsetInitial$ and $\offsetFinal$ are the pre-collision and
post-collision particle offsets.  For simplicity, we assume that the
interacting particles undergo symmetric transverse displacements.  The
first term of the integrand on the right-hand side of
Eq.\ \eqref{population-balance equation} corresponds to collisions
pushing a particle into the position $z$ (moving it from
$z+\halff\offsetInitial-\halff\offsetFinal$), and the second term to
collisions displacing a particle from the position $z$.  If the
initial and final offsets are the same,
$\offsetInitial=\offsetFinal$, the first and second terms of the
integrand cancel.  In the full three-dimensional (3D) version, an
additional integration with respect to the lateral offset $\Delta y$
would be present \cite{supplemental}.

\paragraph*{Collision models M2 and M3 --} 
Based on  geometry of binary collisions we introduce a 2D collision
model M2,
\begin{equation}
\newcommand{\separateit}{&\quad&}
  \label{collision model}
  \offsetFinal = 
\left \{ \begin{array}{lcccc}
      -\offsetInitial, 
         \separateit0<&|\offsetInitial|& < \swaprange,\\
      \sign{\offsetInitial}\deff ,  
         \separateit\swaprange < &|\offsetInitial|& < \deff,\\
      \offsetInitial, 
         \separateit\deff<&|\offsetInitial|&<\infty,
\end{array}  \right.
\end{equation}
where $\swaprange$ is the swapping range and $\deff$ collision range.
For a collision parameter $\offsetInitial$ smaller than the swapping
range $\swaprange$, particle swapping leads to the exchange of
transverse particle positions, consistent with the results of our
hydrodynamic calculations illustrated in Figs.\ \ref{trajectory plot
version 2}(c) and \ref{estability}(a).  Hence, the two terms of the
integrand in Eq.\ \eqref{population-balance equation} cancel each
other.  For the collision parameter in the range $\swaprange <
|\offsetInitial| < \deff$, particle interaction results in finite
particle displacements.

In M3, our 3D model \cite{supplemental}, particles on collisional
trajectories behave as hard spheres without HI.  In the swapping
trajectory region $|\offsetInitial|<\swaprange$ the particles exchange
their transverse positions $z$ (as in model M2).

\paragraph*{Small perturbation analysis --}
We show that the existence of the swapping region
$|\offsetInitial|<\swaprange$ results in a layering instability,
provided that the swapping ratio
\begin{equation}
\label{swapping ratio parameter}
\swappingRatio=\swaprange/\deff
\end{equation}
is sufficiently large.  The conditions for a uniform particle
distribution $n_0$ to become unstable can be derived by analyzing
small harmonic perturbations $\density(z;t)=n_0+n_1(t)
e^{\mathrm{i}kz}$.  For the translationally invariant population-balance
models M2 and M3, such Fourier modes evolve exponentially,
$n_1(t)=n_1(0) e^{\alpha t}$.

Figure \ref{estability}(b) shows that the growth rate $\alpha(k)$ has
a peak around $k \approx 2\pi/\deff$.  Since the peak value becomes
positive at a critical swapping ratio
$\swappingRatio=\swappingRatioCrit$, the uniform particle distribution
is unstable to small perturbations for
$\swappingRatio>\swappingRatioCrit$, leading to formation of particle
layers.  Fig.\ \ref{estability}(b) depicts model M2, but we find that
M3 yields a similar behavior.

Figure \ref{estability}(c) shows the peak values of $\alpha$ plotted
vs. the swapping ratio $\lambda$ (both for M2 and M3).  For M2 the
critical swapping ratio is $\swappingRatioCrit=0.24$, and for M3 we
obtain $\swappingRatioCrit=0.213$.  Our \BMCHI\ simulations yield the
instability in a similar parameter range.

Based on the dynamics of pair collisions (with HI), the swapping ratio
\eqref{swapping ratio parameter} can be controlled in two ways.
First, the swapping range $\swaprange$ can be changed by varying the
wall separation $H$. For example, for a particle pair in the middle of
the channel, the dimensionless range $\swaprange/d$ varies between
0.27 for $H/d=5$ and 0.12 for $H/d=40$
\cite{Zurita_Gotor-Blawzdziewicz-Wajnryb:2007b}. Second, the swapping
ratio can also be controlled by changing particle roughness, because
collision range $\deff$ diminishes with the decreasing roughness
amplitude \cite{daCunha-Hinch:1996}, as shown in Fig.\
\ref{estability}(a).  We thus predict that by decreasing the magnitude
of the particle roughness, we can induce formation of a long-range
order in a confined suspension under shear.  This prediction is
confirmed by our \BMCHI\ simulations (cf., Fig.\ \ref{H15-dens version
2}, where $\swappingRatio=0.163$ for $\epsilon/d=0.25$ and
$\swappingRatio=0.23$ for $\epsilon/d=0.064$, based on the mid-channel
swapping range $\swaprange/d=0.189$).

\paragraph*{Near-wall microstructure --} 

Further predictions regarding the wall-induced suspension ordering can
be obtained when geometrical constraints that disallow particle--wall
overlaps are introduced into the population-balance model
\refeq{population-balance equation}.  Figures \ref{near-wall behavior
  stable} and \ref{near-wall behavior unstable} show our results for
model M3 with such constraints.  We consider a system with a large
wall separation $H/\deff=60$ to emphasize two important regimes of
suspension behavior: subcritical (Fig.\ \ref{near-wall behavior
  stable}) and supercritical (Fig.\ \ref{near-wall behavior
  unstable}).

In both regimes we observe formation of a wall-induced layered order,
but the evolution and extent of the layered microstructure is
regime-specific. For subcritical values of the swapping ratio
\eqref{swapping ratio parameter}, the microstructure appears only in
the near-wall domain, whereas for supercritical values the layered structure
propagates from the wall into the bulk of the suspension.

\paragraph*{Subcritical regime --}

Figure \ref{near-wall behavior stable}(a) shows the subcritical
microstructure for $\swappingRatio=0$ (no swapping) and
$\swappingRatio=0.1$ (significant swapping).  In both cases several
particle layers form near the wall due to the excluded-volume effect.
The correlation length $\correlationL$ of this local layered structure
is insensitive to the value of $\swappingRatio$ in the whole
subcritical regime [cf., Fig.\ \ref{near-wall behavior stable}(b)].
However, for $\swappingRatio=0$ the layers are not sharply defined,
with many particles in the valleys between the density peaks. For
$\swappingRatio=0.1$ the layers are much sharper.  Thus the ST
mechanism influences the form but not the extent of the subcritical
microstructure.

\paragraph*{Supercritical regime --}

In the supercritical regime (cf.\ Fig.\ \ref{near-wall behavior
  unstable}) the wall-induced microstructure propagates from the wall
  with velocity $\vFront$, forming a growing number of particle
  layers, as shown in Fig.\ \ref{near-wall behavior unstable}(a).  In
  a fully developed microstructure the layers are well separated, and
  particle jumps between layers are rare [cf., Fig.\ \ref{trajectory
  plot version 2}(d)].

The propagation velocity $\vFront$ of the layered structure in this
regime [cf. Fig.\ \ref{near-wall behavior unstable}(b)] is an
increasing function of $\swappingRatio$, with two distinct dynamical
domains.  For $\swappingRatio>\swappingRatioCrit$ (suspension unstable
to small perturbations), $\vFront$ rapidly increases with the swapping
ratio, showing a power law critical behavior
$\vFront\sim(\swappingRatio-\swappingRatioCrit)^{1/2}$ [cf. Fig.\
\ref{near-wall behavior unstable}(c)].  In the domain
$\swappingRatioWall<\swappingRatio<\swappingRatioCrit$ the suspension
is stable to small perturbations, but unstable to the {\it large
perturbation} caused by the wall.  In this regime, the data can be
well fitted to the power law
$\vFront\sim(\swappingRatio-\swappingRatioWall)^\beta$ with
$\swappingRatioWall\approx0.11$ and $\beta\approx4$ [cf. Fig.\
\ref{near-wall behavior unstable}(d)].  Due to the large value of the
exponent $\beta$, the propagation velocity $\vFront$ is practically
zero for $\swappingRatio\lesssim0.16$.

The layer separation in the propagating microstructural fronts may be
incommensurate with the wall separation. In such cases the two
mid-channel layers merge, as depicted in Fig.\
\ref{commensurability}(a).  After such suspension restructuring, the
width of the density peaks significantly decreases.  Note that the
initial number of layers agrees fairly well with our prediction based
on the wavelength of the most unstable Fourier mode [cf.  Fig.\
\ref{commensurability}(b)].

\paragraph{Conclusions --} 

The ST mechanism significantly alters
the suspension microstructure in a confined Couette flow by inducing
layering.  Previously (solving a long-standing paradox
\cite{Zarraga-Leighton:2002}),
 we have demonstrated that swapping trajectories cause the anomalous
enhancement of hydrodynamic diffusion, and that they stabilize particle
chains in microfluidic channels
\cite{Zurita_Gotor-Blawzdziewicz-Wajnryb:2007b}.  The layering
behavior identified and analyzed here adds to the growing evidence
that a seemingly subtle ST effect has far-reaching consequences in
confined particulate flows.

Our direct numerical simulations indicate that swapping trajectories
significantly influence a variety of confined systems, including
suspensions at higher concentrations \cite{supplemental} and
flow-driven particle monolayers described in
\cite{Blawzdziewicz-Goodman-Khurana-Wajnryb-Young:2010}.  
Since such monolayers (in the disordered state) have a similar
structure to the layers observed experimentally in
\cite{Cohen-Cheng:2011}, we conclude that the ST effect has
significant implications for suspension dynamics.  In suspensions with
nonzero inertial forces
\cite{Mikulencak-Morris:2004} 
and in viscoelastic fluids, swapping (reversing) trajectories occur
even without confinement, so the ST mechanism is likely to affect
suspension microstructure also in such flows.

Finally, we note that the suspension behavior described in our study
significantly differs from the dynamics of a dilute gas. In a gas,
binary collisions always produce a homogeneous equilibrium state,
according to Boltzmann's H-theorem.  By analogy, it is assumed that
binary collisions in suspension flows always lead to a diffusive
behavior that results in relaxation of density
fluctuations.  Here we report that {\it uncorrelated binary
collisions} produce inhomogeneous layered microstructure.

\acknowledgments{We acknowledge financial support by NSF grant CBET
  1059745 (JB), Ministerio de Innovaci\'on y Ciencia grant
  RYC-2008-03650 (MZG), and Polish Ministry of Science and Higher
  Education Grant No. N N501 156538 (EW).}

\end{document}